# DATA ACCESS — EXPERIENCES IMPLEMENTING AN OBJECT ORIENTED LIBRARY ON VARIOUS PLATFORMS[*]

R. Lange, BESSY, 12489 Berlin, Germany
J. Hill, LANL, Los Alamos, NM 87545, USA

## Abstract

Data Access will be the next generation data abstraction layer for EPICS [1]. Its implementation in C++ brought up a number of issues that are related to object oriented technology's impact on CPU and memory usage. What is gained by the new abstract interface? What is the price that has to be paid for these gains? What compromises seem applicable and affordable? This paper discusses tests that have been made about performance and memory usage as well as the different measures that have been taken to optimize the situation.

## 1 DATA ACCESS

In the next generation of EPICS a redesigned data abstraction layer will replace the existing data container library GDD [2]. An object-oriented library called Data Access will provide this new interface. The background and key design objectives of this approach are described in detail in another contribution to this conference [1].

### 1.1 Features

The first and maybe main application for this library will be its use by the EPICS network protocol Channel Access [3]. In this context Data Access offers a number of key advantages over the existing data conversion library:

- **Extensibility.** Applications (both clients and servers) may define new data container structures that Channel Access will transport across a network.
- **Range checks.** For a set of basic data types and structures thereof conversion routines are provided that include checking data validity.
- **Type safety.** The library interface uses the features that C++ provides for compile time type checking (e.g. overloaded functions for all basic data types).
- **Multi-dimensional arrays.** Arrays of arbitrary size and number of dimensions are supported. Methods for extraction of sub-arrays are provided that include boundary checking. Array data may be read in arbitrary length chunks to allow using buffers of any size for further processing and network transport.
- **Improved conversion table design.** In EPICS the data conversion jump table was originally coded explicitly by hand. With GDD it was program-generated, and with Data Access it is produced by a comparatively compact set of C++ templates.

### 1.2 Target Environments

Using Data Access as the data abstraction layer for Channel Access requires the library to be ported to a large number of target systems with different demands.

Toolkit components running on workstations must cope with a range of available compilers, but code size is less important because old workstations are routinely taken out of service. In contrast, legacy embedded platforms without benefit of virtual memory, such as Motorola 68k based EPICS input output controllers, are not routinely upgraded. These computers place stringent demands on code size and performance.

The C++ compilers for the different target platforms are either rapidly evolving or frozen to a certain version of the embedded real time operating system. Each compiler implements slightly different subsets of the C++ standard. Unfortunately, certain advanced features of C++ (name spaces, local template class members) had to be completely avoided or replaced by workarounds until all compilers support these constructs.

### 1.3 Test Bed

The first implementation was compiled and tested on three platforms using three different compilers:
- Pentium PC running Linux – G++
- Sun Ultra-30 running Solaris – Sun WSpro
- Sun Ultra-30 running Solaris – G++
- Pentium PC running Windows – Microsoft C++
- Motorola 68k running vxWorks – G++

The benchmarks were generated with a small test application using some classes that resemble "typical" containers holding the value, alarm state and timestamp properties. Conversion performance was measured by assigning between containers of different data types. Processor cache related influences have been taken into account by testing with different size arrays of containers. Nevertheless, the results can only provide a first impression of the library's behavior. Further tests on a wider set of platforms with mature compilers are necessary.

[*] Work supported by the Bundesministerium für Bildung, Wissenschaft, Forschung und Technologie (BMBF), the Land Berlin and the Office of Energy Research, Basic Energy Science of the US Department of Energy.

## 2 PERFORMANCE

The data conversion routines will be used multiple times in each transaction between Channel Access and the data on both server and client side. It is obvious that the run-time efficiency of these conversions is crucial for the overall system performance.

### 2.1 Design

One important performance aspect directly influenced the interface design: Modeling the new interface after the existing GDD classes would have led to the accessing methods being implemented as virtual functions. Having every access go through the virtual function table of the data class was found to be too big a performance hit. The current implementation features a callback mechanism, where the user data class calls back into a library-provided adaptor for each container element. This mechanism might appear to be more complicated to use, but it is more efficient.

### 2.2 Observations

The performance highly depends on the machine and the compiler used to generate the library. Discussing detailed performance numbers for the different machines and compilers would certainly go beyond the scope of this paper, but a few statements are safe to make:
- Numeric data container assignment takes about 1 µs on the Pentium/GNU and about 4 µs on SPARC/WSpro.
- On SPARC, the WSpro compiled code takes about 50 % more time when one of the containers' data is unsigned and the other's is signed.
- Conversions between numerical and string data take about 10 times longer than between numerical types — with the exception of taking only about 3 times longer on SPARC/GNU.
- Assignment of integer arrays adds between 0.02 µs and 0.1 µs per element depending on the array size and structure and 0.6 µs per chunk on the Pentium/GNU, 4 times as much on the SPARC/WSpro.

### 2.3 Further Improvements

For contiguous array and sub-array copies of the same data type a specialization was introduced that uses the C runtime function memcpy() instead of element-by-element assignment. This increases the performance in these special but common cases by a factor of 4.

## 3 OBJECT SIZE

With legacy embedded systems the object size of the library is an important issue. Several test series have been made and measures have been taken to reduce the library's object size while retaining the advantages and features shown above.

In this paper's context all size numbers are shown for the converter function classes on the Pentium/Linux/GNU platform. Generally, the same object files are twice as large for the SPARC and half as large for the Motorola 68k target, which reflects the different processor architectures.

The existing C conversion library — with less functionality than the new interface — has an object size of 37 KB.

Despite following Scott Meyer's rules on effective C++ programming [4,5] as close as possible, a point was reached during the implementation of Data Access when the G++ compiler on Sun wasn't able to compile the library anymore. The compiler process grew to 350 MB resident process size and had to be stopped after an hour without any results. The Sun WSpro managed to compile the code, but took almost 40 minutes. In contrast, Linux and Windows compilers compiled the code without unreasonable delay. At this point the size of the basic conversion template classes for array and scalar data had grown to 8 MB and 5MB — a total of 13 MB object code generated from less than 1000 lines of source. That certainly wasn't tolerable.

A number of measures have been taken to improve the situation. Table 1 shows an overview of the resulting object sizes.

### 3.1 Optimization and Debug Information

Templates and inline function calls may create an enormous amount of debugging information. Setting reasonable compiler switches to avoid generation of debug information and to enable a suitable level of optimization shrunk down the objects to 1 MB and 360 KB respectively.

### 3.2 Use of Templates

Data Access uses templates extensively. A fairly large number of classes are templates with one type parameter: the user data type (in this case from a set of $n=15$ user data types). The central data conversion functions naturally are templates with two formal type parameters: the source and the destination data type. Therefore within a universal library, these classes get instantiated $n \times n = 225$ times. So every line of code or data within these classes increases the object size by 225 times its size.

### 3.3 Exceptions

When using exceptions, additional object code is generated by the compiler [6]. This code sums up to a few hundred bytes per throw(). Simply putting the calls to throw() as static members into an external class reduced the code size further down to 375 KB and 200 KB respectively.

### 3.4 Inline Declarations and Repeated Code

Inline functions increase performance by avoiding a function call. Their code is repeated at every use, which is usually tolerable, but within templates — multiplied by the number of instantiations — they must be used judiciously. The code was changed to avoid calling inline functions within templates where this was inappropriate.

Originally the recursive algorithm to copy arrays of arbitrary size and number of dimensions was placed within the template conversion class. The class design was changed to move this functionality out of this class into the user's class.

These measures reduced the object size down to 195 KB (in both cases).

### 3.5 Implicit Conversion

Another step uses the fact that C++ implicitly converts function arguments if there is no precision loss (i.e. from smaller to wider formats). Therefore some template functions have to be instantiated only for a few source types — the compiler will promote the arguments to a canonical type. This change reduced the object sizes to the current values of 193 KB and 132 KB respectively.

Table 1: Object Sizes for Conversion Routines

| Status | Array | Scalar |
|---|---|---|
| Initial | 8 MB | 5 MB |
| No debug info, optimized | 1 MB | 360 KB |
| Without throw() | 375 KB | 200 KB |
| Without archive copy code and inline definitions | 195 KB | 195 KB |
| With implicit conversion | 193 KB | 132 KB |

## 4 CONCLUSIONS

Data Access, the next generation data abstraction layer for EPICS, provides a number of important advantages over the existing interfaces. Its implementation in C++ makes extensive use of the powerful constructs the language provides, thereby considerably reducing the size of the source code.

The existing C++ compilers do not fully implement the C++ standard — thus writing portable code can be tedious. Workarounds must be introduced for temporarily missing language features.

The compilers also show significant differences in compile time and efficiency. The resulting object code covers a wide size and performance range. Some of the native compilers take a surprisingly long time to generate slow and bloated code.

Using C++ and object oriented technologies in performance critical low-level libraries introduces a number of potential problems and difficulties that one might not expect. It is necessary to design and implement with caution: powerful language constructs can multiply side effects that may outnumber the advantages gained. The current C++ compilers make it impossible to simply implement a clear straightforward design. The programmer has to take into account extensive details about each compiler's implementation of certain language features, if the code needs to be portable.

Nevertheless, with all the optimizations that have been applied, we are certain that the remaining performance and size overhead will be neglectable compared to the benefits available with the improved interface.

## REFERENCES

[1] J. Hill, R. Lange: "Next Generation EPICS Interface to Abstract Data", ICALEPCS 2001 Conference, San Jose, USA.

[2] J. Kowalkowski: "General Data Descriptor Library User's Guide and Reference Manual", 1996, APS, Argonne, USA[1].

[3] J. Hill: "EPICS R3.14 Channel Access Reference Manual", 2001, LANL, Los Alamos, USA.

[4] S. Meyers: "Effective C++", second edition, Addison Wesley Longman Inc., 1998.

[5] S. Meyers: "More Effective C++", Addison Wesley Longman Inc., 1996.

[6] T. Cargill: "Exception Handling: A False Sense of Security", first published in: C++ Report, November-December 1994, included in the CD edition of [4,5].

---

[1] http://www.aps.anl.gov/asd/controls/epics/EpicsDocumentation/EpicsGeneral/gdd.html